\documentclass{emulateapj}

\def\lsim{\lower.5ex\hbox{$\; \buildrel < \over \sim \;$}}
\def\gsim{\lower.5ex\hbox{$\; \buildrel > \over \sim \;$}}

\def\ch{\lower-0.55ex\hbox{--}\kern-0.55em{\lower0.15ex\hbox{$h$}}}
\def\lh{\lower-0.55ex\hbox{--}\kern-0.55em{\lower0.15ex\hbox{$\lambda$}}}
\def\mc{meridional circulation}
\def\bl{Babcock--Leighton}
\def\Rs{R_{\odot}}

\def\lsim{\lower.5ex\hbox{$\; \buildrel < \over \sim \;$}}
\def\gsim{\lower.5ex\hbox{$\; \buildrel > \over \sim \;$}}
\def\ch{\lower-0.55ex\hbox{--}\kern-0.55em{\lower0.15ex\hbox{$h$}}}
\def\lh{\lower-0.55ex\hbox{--}\kern-0.55em{\lower0.15ex\hbox{$\lambda$}}}
\shorttitle{IMPORTANCE OF MERIDIONAL CIRCULATION IN FLUX TRANSPORT DYNAMO}
\shortauthors{Karak}

\begin{document}
\title{IMPORTANCE OF MERIDIONAL CIRCULATION IN FLUX TRANSPORT DYNAMO:\\
THE POSSIBILITY OF A MAUNDER-LIKE GRAND MINIMUM}
\author{BIDYA BINAY KARAK}
\affil{Department of Physics, Indian Institute of Science, Bangalore 560012, India}
\altaffiltext{}{bidya$\_$karak@physics.iisc.ernet.in}

\begin{abstract}
Meridional circulation is an important ingredient in 
flux transport dynamo models. We have studied 
its importance on the period, the amplitude of the solar cycle, 
and also in producing Maunder-like grand minima in these models. 
First, we model the periods of 
the last $23$ sunspot cycles by varying the meridional circulation speed. 
If the dynamo is in a diffusion-dominated regime, then we find that most of 
the cycle amplitudes also get modeled up to some extent when we model 
the periods. Next, we propose that at the beginning of the Maunder minimum 
the amplitude of meridional circulation dropped to a low value and then 
after a few years it increased again. Several independent 
studies also favor this assumption. With this assumption, a 
diffusion-dominated dynamo is able to reproduce many important
features of the Maunder minimum remarkably well. 
If the dynamo is in a diffusion-dominated regime, then a slower meridional 
circulation means that the poloidal field gets more time to diffuse 
during its transport through the convection zone, making the dynamo weaker. 
This consequence helps to model both the cycle amplitudes and the Maunder-like 
minima. We, however, fail to reproduce these results if the dynamo is in 
an advection-dominated regime.
\end{abstract}
\section{Introduction}
An important aspect of solar activity are the grand minima periods during which 
the activity level is strongly reduced. The best example of this is the 
Maunder minimum during 1645--1715 (Eddy 1976; Ribes \& Nesme-Ribes 1993). 
It is important not only to the Solar Physics Community but also to the Space Weather and Earth Climate 
Community as it may be correlated to the Little Ice Age. It is not an artifact of few observations, but a 
real phenomenon \cite{hoyt}. There are evidences that solar-like stars also show Maunder-like minima 
\cite{baliu}. The Maunder minimum episode may be divided into two periods: one from 
1645 to 1670 and another from 1670 to 1715. During the first period, the sunspot number was almost zero 
in both the hemispheres, whereas during the second period, a few sunspots appeared in the southern 
hemisphere. So there was a strong north--south asymmetry of sunspot number in the last phase of the 
Maunder minimum (Ribes \& Nesme-Ribes 1993; Sokoloff \& Nesme-Ribes 1994). Another important feature 
is the sudden initiation, but slow recovery \cite{usos00}.
From the study of the cosmogenic isotope $^{10}$Be in polar ice core, 
Beer et al. (1998) suggested that the usual 11~year period of solar activity (Schwabe cycle) 
continued during the Maunder minimum, although the overall activity level was weaker. 
However, from the study of $^{14}$C data in tree rings, Miyahara et al. (2004) 
found the same cyclic behavior but with a period of 13--15 years instead of the regular 11~year 
period (see also Miyahara et al. 2010).

Flux transport dynamo models developed by several authors (Wang et al. 1991; 
Choudhuri et al. 1995; Durney 1995; Dikpati \& Charbonneau 1999; 
K\"uker et al. 2001; Nandy \& Choudhuri 2002; Chatterjee et al. 2004; Guerrero \& Mu\~noz 2004; 
Mu\~noz-Jaramillo et al. 2009; Hotta \& Yokoyama 2010a) are the most 
promising models for studying the solar cycle. 
It explains many important features of the regular 
solar cycle including 11~year periodicity, equatorial migration of the 
sunspot belt, poleward migration of large-scale field and parity, etc. 
It also reproduces many irregular features of the solar cycle 
(Charbonneau \& Dikpati 2000; Choudhuri \& Karak 2009; 
Karak \& Choudhuri 2010a, 2010b). The flux transport dynamo model 
consists of the following processes. 
A strong toroidal component of the magnetic field is generated 
in the tachocline due to the strong differential rotation \cite{parker1955}. 
When this toroidal magnetic field rises above the stable layer, 
the magnetic buoyancy acts on it causing it to rise to the 
surface to form sunspots. These sunspots decay to give rise to 
the poloidal component of the magnetic field through the Babcock--Leighton mechanism 
(Babcock 1961; Leighton 1969), which then gets advected (also diffused) 
to the pole by the meridional circulation. 
This field ultimately reaches the tachocline where it again 
gets stretched to give rise to the toroidal field and the cycle
completes. In this process, the poloidal field is produced from the
decay of the tilted bipolar sunspots. These tilts are due to the Coriolis force acting
on the rising flux tubes (D'Silva \& Choudhuri 1993). In addition, when these flux tubes rise
through the convection zone, they are subjected to the convective buffeting 
(Longcope \& Choudhuri 2002).
This gives rise to a scatter in the tilt angles around the mean given by Joy's law
\cite{wang}. We therefore believe that the Babcock--Leighton process is not a 
deterministic process. As a result, the polar field changes randomly as its
production is governed by the random process \cite{ccj}. Following this idea, 
Choudhuri \& Karak (2009) have proposed that at the beginning of the 
Maunder minimum, the polar field fell to a very low value. 
With this assumption, they have succeeded in reproducing a Maunder 
minimum using a flux transport dynamo model. 
Some authors (Choudhuri 1992; G\'omez \& Mininni 2006; Wilmot-Smith et al. 2005; 
Brandenburg \& Spiegel 2008; Moss et al. 2008; Usoskin et al. 2009) 
are also able to reproduce Maunder-like grand minima by introducing stochastic 
fluctuations or nonlinearity in dynamo parameters. By introducing stochastic fluctuations 
in the flux transport dynamo simulations, Charbonneau et al. (2004) are 
able to reproduce intermittencies which resemble the Maunder minimum. 
On the other hand, Beer et al. (1998) proposed the nonlinearity to be the cause 
of such irregularities of the solar cycle.

Another vital source of irregularity in the flux transport dynamo model is the meridional 
circulation which plays a major role in transporting the magnetic fields. 
It determines not only the period 
but also the amplitude of the solar cycle (Wang et al. 1991; Dikpati \& Charbonneau 1999; 
Hathaway et al. 2003; Yeates et al. 2008). 
It also varies stochastically with time, giving randomness in the solar activity. Recently 
several authors (Hathaway 1996; Haber et al. 2002; Basu \& Antia 2003; 
Gizon \& Rempel 2008; Hathaway \& Rightmire 2010) have 
reported its value in the upper 
part of the convection zone and found strong temporal variation of its 
amplitude. Javaraiah \& Ulrich (2006) studied group sunspot data during cycles 12--23 
and found a cycle-to-cycle variation of mean meridional motion of sunspot groups (a proxy 
of the meridional flow). Several other indirect studies (Wang et al. 2002; Hathaway et al. 
2003; Georgieva \& Kirov 2010) have found the evidences of the fluctuations of the meridional 
circulation. Moreover, using a low order model constructed 
from the sunspot number as the proxy of the toroidal field, Passos \& Lopes (2008) have 
suggested that the last $11$ magnetic cycles can be modeled using variable meridional 
circulation speed only. Using similar a model, Passos \& Lopes (2009) 
have also concluded that the stochastic fluctuations in the $\alpha$-effect cannot 
trigger a grand minimum, rather they argued that the strong decrease of meridional 
circulation can do so.

The motivation of this work is to find out and understand the influence 
of the meridional circulation on both cycle-to-cycle amplitude variation 
of the the solar cycle and producing grand minima. Therefore, first, we try to model the periods of 
the last $23$ sunspot cycles just by varying the amplitude of 
\mc\ to see its effect on the amplitude of the solar cycle. 
We have shown the results from 
a diffusion-dominated flux transport dynamo model (e.g., Chatterjee et al. 2004). By the 
term ``diffusion-dominated'' we mean that the diffusivity of the poloidal field in the whole 
convection zone is high enough ($\sim 10^{12}$--$10^{13}$ cm$^2$ s$^{-1}$; see also 
Jiang et al. 2007 and Yeates et al. 2008 for details). However, we discuss the 
result from advection-dominated flux transport dynamo model (e.g., 
Dikpati \& Charbonneau 1999) as well. By the term ``advection-dominated'' we mean 
that the diffusivity of the poloidal field in the whole convection zone is low ($\sim 
10^{10}$--$10^{11}$ cm$^2$ s$^{-1}$). In this model, therefore, the diffusivity 
is not so important in transporting the fields. In a high-diffusivity model, 
we have found that most of the cycle amplitudes get modeled up to some extent when 
we try to model the periods of the last $23$ cycles just by varying the 
amplitude of \mc\ only. Therefore, we conclude that a major part of the 
fluctuations of the amplitude of the solar cycle may come from the fluctuations 
in meridional circulation. However, we do get a completely different result, 
if the dynamo is in the advection-dominated regime. The physics of getting 
these two completely different results from these two models can be 
understood based on the Yeates et al. (2008) study what we mention in the 
appropriate place.

Next, we also study the importance of meridional circulation in producing 
a Maunder-like grand minimum in both the high-diffusivity model and the 
low-diffusivity model. We propose that at the beginning of the Maunder minimum, 
the amplitude of meridional circulation dropped to a low value and then 
after a few years it increased again to the original value. With this 
assumption, we have checked the possibility of producing (or not producing) 
a Maunder-like grand minimum in both the models. In the next step, 
we have also included the fluctuations of polar field to capture the 
fluctuations in the \bl\ process along with 
that of \mc\ to model such kind of grand minimum.
\section{MODEL}
The evolution of poloidal and toroidal components of the magnetic field 
in the flux transport dynamo models is governed, respectively, by the 
following two equations:
\begin{equation}
\frac{\partial A}{\partial t} + \frac{1}{s}({\bf v}.\nabla)(s A)
= \eta_{p} \left( \nabla^2 - \frac{1}{s^2} \right) A + \alpha B,
\end{equation}
\begin{eqnarray}
\frac{\partial B}{\partial t}
+ \frac{1}{r} \left[ \frac{\partial}{\partial r}
(r v_r B) + \frac{\partial}{\partial \theta}(v_{\theta} B) \right]
= \eta_{t} \left( \nabla^2 - \frac{1}{s^2} \right) B \nonumber \\
+ s({\bf B}_p.{\bf \nabla})\Omega + \frac{1}{r}\frac{d\eta_t}{dr}\frac{\partial{(rB)}}{\partial{r}},~~~~~~~~~~~~~~~~~~~~~~~~~~~~~
\end{eqnarray}\\
where $A(r, \theta)$ and $B (r, \theta)$, respectively, correspond to the 
poloidal and toroidal components and $s = r \sin \theta$.  
Here ${\bf v}$ is the velocity of the meridional flow, $\alpha$ is the coefficient 
which describes the generation of the poloidal field near the solar surface from 
the decay of bipolar sunspots, $\Omega$ is the internal angular velocity of 
the Sun, and $\eta_p$, $\eta_t$ are the turbulent diffusivities for the poloidal 
and toroidal fields.

To study the evolution of the magnetic fields, we have to 
solve above two equations with the given parameters. 
Recently this problem has been extensively studied by 
two different groups (Dikpati \& Charbonneau 1999 and 
Chatterjee et al. 2004). Dikpati \& Charbonneau (1999) 
(also Dikpati et al. 2004) use a very low value 
($10^{10}$--$10^{11}$ cm$^2$ s$^{-1}$) of the diffusivity 
of the poloidal field in the convection zone, whereas 
Chatterjee et al. (2004) (also Jiang et al. 2007) use 
very high value ($10^{12}$--$10^{13}$ cm$^2$~s$^{-1}$). 
Although both the models can explain
some important features of the solar cycle (e.g., 11~year periodicity,
equatorward propagation of the sunspot belt) (Dikpati \& Charbonneau 1999;
Chatterjee et al. 2004), in many cases, they give completely 
different results (Chatterjee et al. 2004; Jiang et al. 2007; 
Yeates et al. 2008; Karak \& Choudhuri 2010b). Therefore, 
we are curious to see the importance of meridional circulation 
in modeling both the solar cycle and the Maunder-like grand minimum in these 
two models separately. To do this, we first use the high-diffusivity model 
of Chatterjee et al. (2004) with the same profiles of 
$\alpha$ coefficient, differential rotation, meridional circulation 
and diffusivity as given in the ``standard model'' of Chatterjee et al. (2004). We do not 
repeat them here. However, Karak \& Choudhuri (2010b) have changed 
some parameters of this model which are listed in Table~2 of their paper. 
In the present work, we use these parameters except we take $\Gamma$ = 
$3.025\times10^{8}$ m and the surface value of \mc\ $v_0$ = $22$ m~s$^{-1}$. 
With these parameters the period of the solar cycle comes out to be around 11~year. 
In all the analyses, only the amplitude of \mc\ is varied while all other 
parameters of the model are held fixed.
A related question at this stage may arise that any change in 
the \mc\ may be accompanied by the change in the differential rotation. 
Since the present understanding of these two is very primitive and 
our aim is to understand the effect of the fluctuation of \mc\ on the 
solar cycle, we have taken differential rotation to be constant in all the analyses.

For the low-diffusivity calculations, if we decrease the diffusivity 
to a very low value ($\sim 10^{10}$ cm$^2$ s$^{-1}$) in the previous 
model, we do not get an oscillatory solution. 
Therefore, we have tried to use the Dikpati et al. (2004) 
model, but we (and also Jiang et al. 2007; Hotta \& Yokoyama 2010a; 
Karak \& Choudhuri 2010b) are not able to reproduce their results. Finally, 
we use the Dikpati \& Charbonneau (1999) model with the same parameters as listed in 
their `reference solution' except for the value of $u_0$. 
To get an 11~year period in their model, we use $u_0$ = $14.5$ m~s$^{-1}$ rather than 
$10$ m~s$^{-1}$ what they have reported in their ``reference solution''. An important 
point to note is that the maximum amplitude of meridional circulation at the 
surface (what Chatterjee et al. 2004 call $v_0$ in their model) is much less 
than this $u_0$.

\section{METHODOLOGY}
\begin{figure*}
\centering
\includegraphics[width=1.00\textwidth]{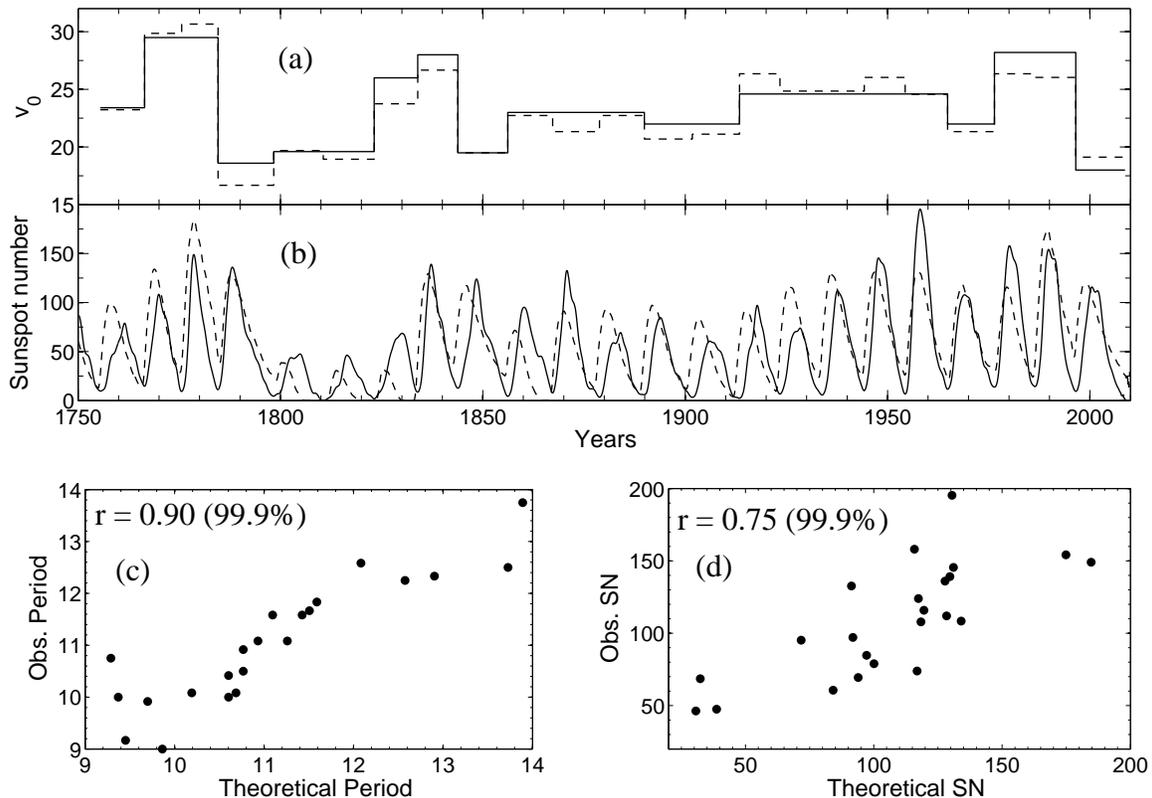}
\caption{(a) Variation of amplitudes of \mc\ $v_0$ (in m~s$^{-1}$) with time 
(in yr). Dashed line is calculated from the relation $T = 97.48v_0^{-0.696}$ 
by putting the periods of the last $23$ cycles data. The solid line is the 
variation of $v_0$ used to match the theoretical periods with the 
observed periods. (b) Variation of theoretical sunspot number (dashed 
line) and observed sunspot number (solid line) with time. Both 
sunspot numbers are smoothed by a Gaussian filter of FWHM = $1$~year. 
In addition, the theoretical sunspot number is scaled by a factor 
to match the observed value. (c) Scatter diagram showing 
the theoretical and observed periods (in yr). 
(d) Scatter diagram showing peak theoretical sunspot number and peak observed 
sunspot number. The linear correlation coefficients and the corresponding 
significance levels are shown in panels (c) and (d).}
\label{fit23}
\end{figure*}

First, we model the last $23$ cycles by fitting the periods with variable 
meridional circulation in a high-diffusivity model. We have done this in 
the following way. It is known that the \mc\ acts as a clock to regulate 
the period of the solar cycle (Wang et al. 1991; Dikpati \& Charbonneau 1999; 
Hathaway et al. 2003). 
Stronger \mc\ gives a shorter period and vice versa. It is found that in the 
flux transport dynamo model, the period of the solar cycle roughly scales as 
the inverse of the \mc\ amplitude. Therefore, we first calculate the 
dependency of period ($T$) on $v_0$ by running the 
model at different values of $v_0$ and we find a relation 
$T = 97.48v_0^{-0.696}$ (when $19\le v_0\le 30$). 
Then from the observed periods of the last $23$ cycles we can calculate the corresponding $v_0$ 
from the above relation. The dashed line in Fig.~\ref{fit23}(a) shows this variation. 
However, these values of $v_0$ do not reproduce the observed periods properly; therefore, we 
fit them by trial and error methods. We did not try to match the periods of each 
cycle accurately which is a little bit difficult. We change $v_0$ abruptly between two cycles 
and not during a cycle. In addition, we do not change $v_0$ if the 
period difference between two successive cycles is less than 
$5\%$ of the average period. The solid line in Fig.~\ref{fit23}(a) shows the variation of $v_0$ 
used to model the periods of the cycles.

\begin{figure*}
\centering
\includegraphics[width=1.00\textwidth]{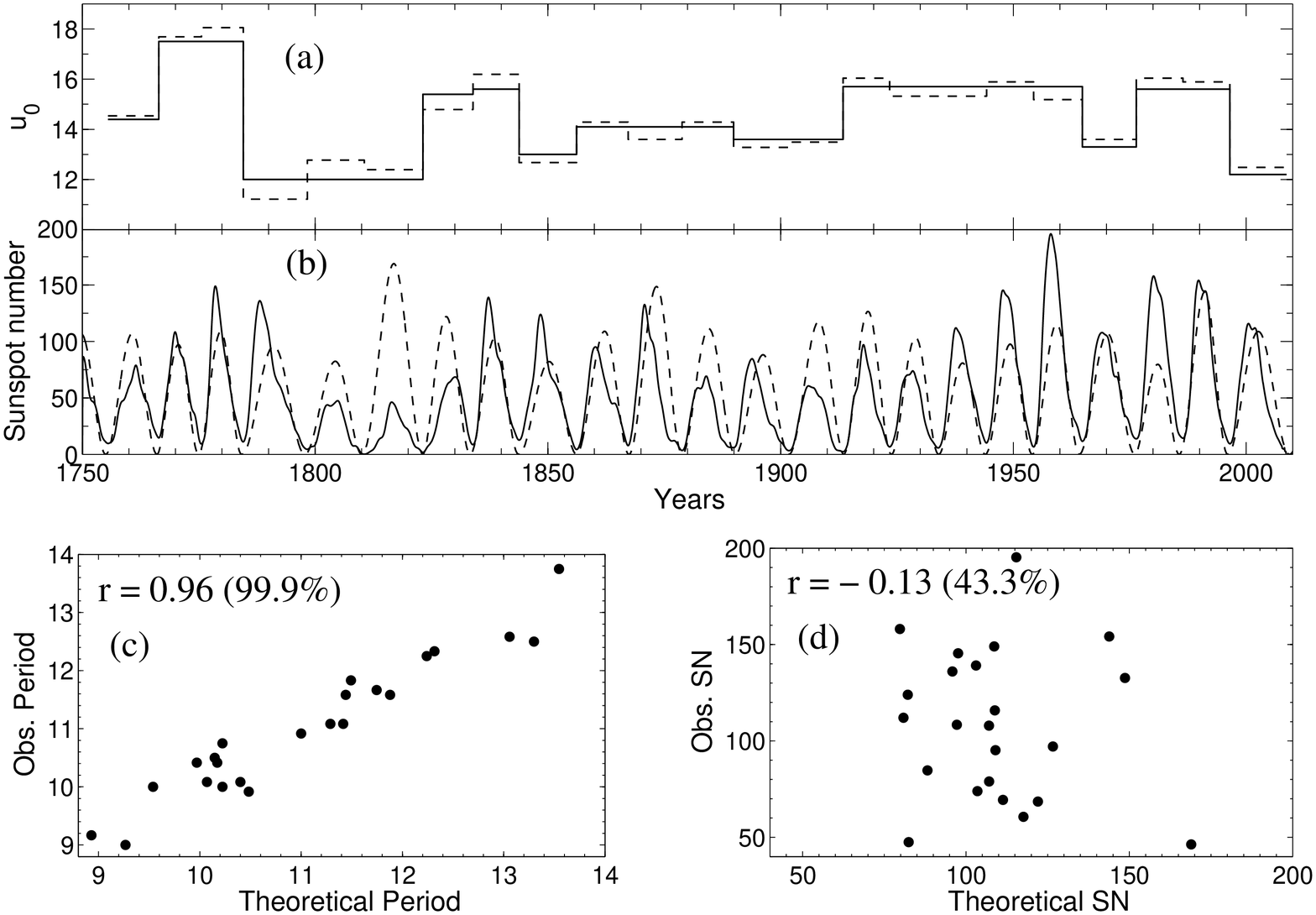}
\caption{Same as Fig.~\ref{fit23}, but from the low-diffusivity model. Here 
the dashed line in panel (a) shows the variation of $u_0$ calculated from Eq.~(12) 
of Dikpati \& Charbonneau (1999) by putting the observed periods of the last $23$ cycles.}
\label{fitlow}
\end{figure*}

Now, we repeat the same analysis in a low-diffusivity model too. 
In this model, the dependency of the period on $u_0$ is given by 
Eq.~(12) of Dikpati \& Charbonneau (1999). From the observed periods 
we calculate $u_0$ according to this relation and we show this 
variation by the dashed line in Fig~\ref{fitlow}(a). Then we 
fit the periods of the last $23$ cycles by trial and error in the same 
way as we have done earlier. 

Next, to reproduce the Maunder minimum, we decrease $v_0$ rapidly to 
a very low value in both the hemispheres. We have done this in the 
decaying phase of the last sunspot cycle before the Maunder minimum. 
Then we keep $v_0$ at a low value for around 1 year and then we again 
increase it slowly to the usual value but at different rates in two 
hemispheres. In northern hemisphere, $v_0$ is increased at a slightly 
lower rate than southern hemisphere. Note that we have varied only $v_0$ 
and no other parameters of the model.

In the next step, we have included the effect of the fluctuations in 
the \bl\ process of generating the poloidal field along with the fluctuations of 
meridional circulation. As the \bl\ process involves randomness, the 
poloidal field at the end of a 
cycle will be different from the average field obtained in the mean 
field model. Choudhuri et al. (2007) have proposed that the 
cumulative effect of the fluctuations of the \bl\ process can be taken 
into account just by multiplying a random factor $\gamma$ with the 
poloidal fields above $0.8 \Rs$ at every minimum of the solar cycle. 
The poloidal fields below $0.8 \Rs$ are left unchanged as it is 
believed that these are from the earlier cycles. 
We follow the same algorithm here. Hence, we decreased the poloidal 
field by a factor $\gamma$ along with $v_0$. We run the model for 
different values of $\gamma$ from 0 to 1 at each value of $v_0$ 
from a very low value to the average value. Then we find out the 
critical values of $v_0$ and the corresponding 
$\gamma$ factor for which we get a Maunder-like minimum. 

We have also checked whether a strong fluctuation of \mc\ can produce 
a Maunder-like grand minimum in the low-diffusivity model.
To do this, we have decreased $u_0$ at the same ratio 
as we have done earlier in the high-diffusivity model. However, 
here we do not include the effect of fluctuations of the polar 
field along with the fluctuations of meridional circulation as we 
have done earlier.

\section{RESULTS}
\subsection{Modeling last $23$ cycles}
First, we discuss the results of fitting the periods of last 
the $23$ cycles by varying $v_0$ in the high-diffusivity model. The 
results are shown in Fig.~\ref{fit23}. It may be noted that 
we have not tried to match the periods exactly. In order to show 
how good the theoretical periods have been matched with that 
of the observed periods, we have shown the scatter 
diagram between them in Fig.~\ref{fit23}(c) (linear corr. coeff. = $0.90$).
The solid line in Fig.~\ref{fit23}(a) shows the variation of $v_0$ 
required to fit the periods of 
the last $23$ cycles. As this $v_0$ determines the period of the solar cycle, 
this may be taken as an indicative of how $v_0$ had varied over the last few centuries. 
It appears from the same figure that during cycles 1--10, the meridional
circulation had large variation with a timescale of around $15$~years. On the other hand, 
during cycles 11--22, the meridional circulation had relatively smaller variation 
with a timescale of around $45$~years. With the limited data we have, we cannot say 
whether the behavior of cycles 1--10 is more typical in the long run or the behavior 
of cycles 11--22 is more typical. Looking at Fig.~\ref{fit23}(a), we can only surmise 
that the meridional circulation probably has long-time variations having large 
coherence time. Indeed, Karak \& Choudhuri (2010b) have also assumed longer coherence 
time of the meridional circulation while studying the Waldmeier effect. 
So just by fitting the periods of the last $23$ 
cycles data we get some idea about both the amplitude variation and the timescale 
of meridional circulation over the last few centuries. Next, in Fig.~\ref{fit23}(b), 
we show the theoretical sunspot series (eruptions) by the dashed line along 
with the observed sunspot series by the solid line. The theoretical 
sunspot series has been multiplied 
by a factor to match the observed value. It is very interesting to see that the amplitude 
of the theoretical sunspot cycle has a similar trend as that of the observed sunspot cycle. 
In order to see the correlation between the amplitudes of peak theoretical sunspot number and 
that of observed sunspot number, we have shown the scatter diagram between 
them in Fig.~\ref{fit23}(d). We have found a significant correlation between these two 
(having linear Pearson's corr. coeff. = $0.75$ with a significance level of $99.9\%$). 
This suggests that a major part of the fluctuations of the amplitude of the solar cycle 
may come from the fluctuations of meridional circulation. This is a very important 
result of this analysis.

It is also interesting to see the results from the low-diffusivity model. 
We show the results of this analysis in Fig~\ref{fitlow}. The solid 
line in Fig~\ref{fitlow}(a) shows the variation of $u_0$ over the 
last $23$ cycles required to fit the periods of the cycles. 
It may be noted that in this model the value of $u_0$ 
required to get an 11 year period is $14.5$ m~s$^{-1}$. 
Although the amplitude of \mc\ at the surface (what in the previous model is 
$v_0$) is different than $u_0$, this $u_0$ variation 
in Fig~\ref{fitlow}(a) is similar to the previous result from 
the high-diffusivity model. Therefore, the low-diffusivity model 
also gives a similar result of the amplitude variation and the 
timescale variation of meridional circulation.
Now let us see how the amplitude of the sunspot cycle responds 
with the amplitude of meridional circulation. In this 
model, the magnetic energy density of the toroidal field ($B^2$) 
at latitude $15$$^{\circ}$ at the base of the convection zone is a 
measure of the sunspot number (Dikpati \& Charbonneau 1999) 
which we also have followed. In Fig.~\ref{fitlow}(b), we have 
shown the theoretical sunspot number by the dashed line along 
with the observed sunspot number by the solid line. We see 
from this figure that the amplitudes 
of this theoretical sunspot number are not matching with 
the observed amplitudes. Rather we have found a weak inverse 
correlation between these two amplitudes (linear corr. coeff. = $-0.13$).

Now we explain the physics behind these two completely different 
results. It is well accepted that the period of the solar cycle is 
strongly determined by the value of $v_0$. This is true in 
both the high-diffusivity model and the low-diffusivity model. However, 
the value of $v_0$ affects the amplitude of the solar cycle differently in these two 
models. This can be understood on the basis of the Yeates et al. (2008) study in 
the following way. We have seen that in these models, the toroidal field is 
generated by the stretching of the poloidal field in the tachocline. The 
production of this toroidal field is more if the poloidal field remains 
in the tachocline for a longer time and vice versa. However, the poloidal field 
diffuses during its transport through the convection zone. As a result, if 
the diffusivity is very high, then much of the poloidal field diffuses away 
and very less amount of it reaches the tachocline to induct the toroidal field. 
Therefore, when we decrease the value of $v_0$ in the high-diffusivity model to match 
the period of a longer cycle, the poloidal field gets more time to diffuse 
during its transport through the convection zone. This ultimately leads 
to a lesser generation of the toroidal field and hence the cycle becomes 
weaker. On the other hand, when we increase
the value of $v_0$ to match the period of a shorter cycle, the poloidal
field does not get much time to diffuse in the convection zone. Hence, it 
effectively gives rise to a stronger toroidal field and the cycle becomes stronger.
Consequently, in the high-diffusivity model, we get weaker 
amplitudes for longer periods (having lower values of meridional circulation) 
and vice versa. This is exactly observed in the observational data (see figure 
1(C) of Charbonneau \& Dikpati 2000). However, this is not the case in the low-diffusivity model. 
This is because in this model the diffusive decay of the fields is not 
much important. As a result, the slower meridional circulation means that 
the poloidal field remains in the tachocline for a longer time and therefore 
it produces more toroidal field, giving rise to a strong cycle. Ultimately, 
we get stronger amplitudes for longer periods (having lower values of meridional 
circulation) and vice versa which is not observed. Therefore, we do not get 
a correct correlation between the amplitudes of theoretical sunspot number and 
that of observed sunspot number when only the meridional circulation is varied.

\subsection{Modeling Maunder minimum}
\begin{figure*}
\centering
\includegraphics[width=1.00\textwidth]{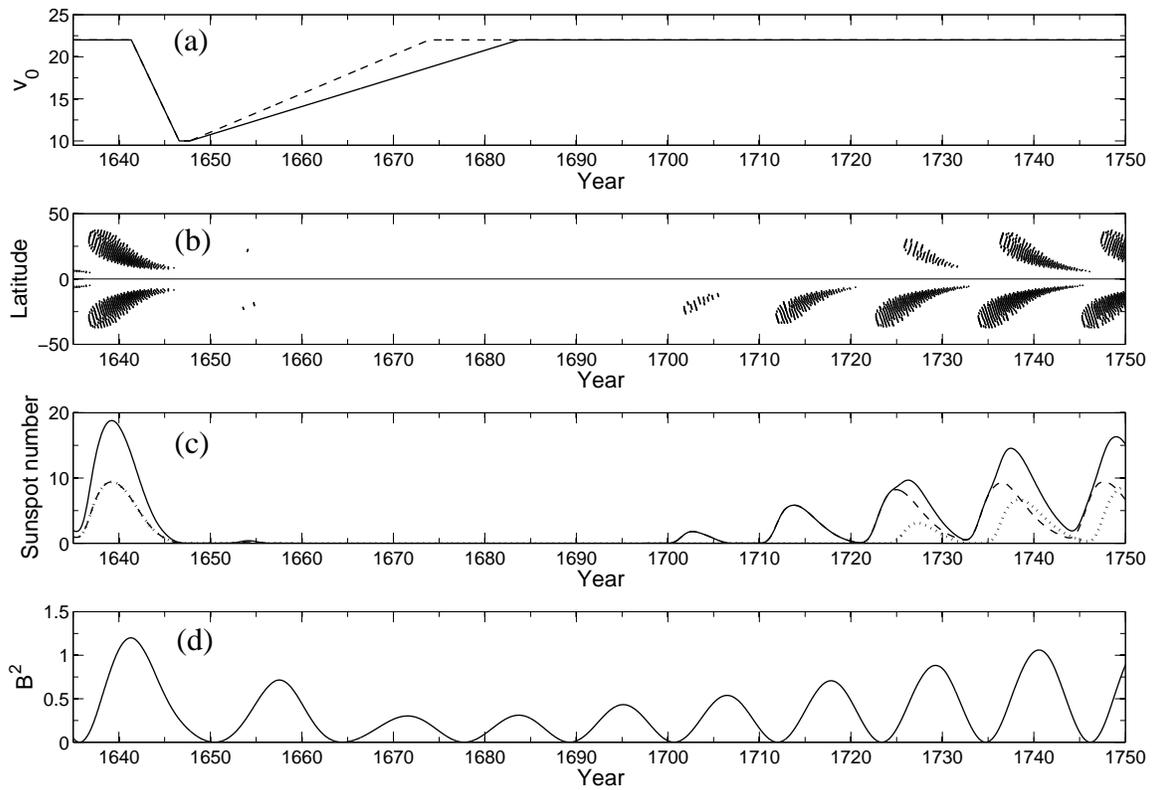}
\caption{Results from the high-diffusivity model. (a) The solid and dashed lines 
show the variations of maximum amplitudes of meridional circulation (in m~s$^{-1}$) in northern and
southern hemispheres with time. (b) The theoretical butterfly diagram. (c) The theoretical sunspot number
(this is obtained by first considering the monthly number of eruptions and then
smoothing it by taking a running average over 5 months iteratively). The dashed
and dotted lines show the sunspot numbers in southern and northern hemispheres, 
whereas the solid line is the total sunspot number. (d) Variation of energy density of the toroidal field 
at latitude 15$^{\circ}$ at the bottom of the convection zone.}
\label{mm}
\end{figure*}

In Fig.~\ref{mm}, we show the theoretical results covering the Maunder
minimum episode from the high-diffusivity model. 
Fig.~\ref{mm}(a), shows the maximum amplitude of meridional 
circulation $v_0$ varied over this period in two hemispheres. 
The key point to note is that in our calculation 
we have increased $v_0$ differently in two 
hemispheres in the recovery phase. In northern hemisphere 
(shown by the solid line in Fig.~\ref{mm}(a)), $v_0$ is increased slightly at a lower rate than 
southern hemisphere (shown by the dashed line). Consideration of this north--south 
asymmetry of \mc\ may be justified. It is unlikely that the 
meridional circulation decreased by the same amount in both hemispheres. 
Even if it did, it is very unlikely that it again increased at the same 
rate. In addition, recent 
observational studies \cite{haber, basu3} of surface \mc\ give 
the evidence of hemispheric asymmetry of \mc\ which can lead to the 
asymmetry in solar activity \cite{dikpati04}. In Fig.~\ref{mm}(b), we 
show the butterfly diagram of sunspot numbers, whereas 
in Fig.~\ref{mm}(c), we show the variation of total sunspot number 
along with the individual sunspot numbers in two hemispheres 
(see the caption of Fig.~\ref{mm}(c)). In order to facilitate comparison 
with observational data, we have taken the beginning of the year to be 1635. 
Note that in Figs.~\ref{mm}(b) and (c), the sunspot number at the beginning 
of the Maunder minimum suddenly falls to zero value, 
whereas in the recovery phase, it increases slowly. In addition, the 
north--south asymmetry of sunspot number observed in the 
last phase of the Maunder minimum is remarkably reproduced. In order to compare 
our results with the observational results, the readers may compare our 
Fig.~\ref{mm}(b) with figure 1(a) of Sokoloff \& Nesme-Ribes (1994) 
and Fig.~\ref{mm}(c) with figure 1 of Usoskin et al. (2000). 
We also show the energy density of the toroidal field calculated 
at latitude 15$^{\circ}$ at the bottom of the convection zone in Fig.~\ref{mm}(d). 
Here, the rapid decrease of the $v_0$ to a critical value ($10$~m~s$^{-1}$) from its 
average value ($22$ m~s$^{-1}$) triggers a Maunder minimum. However, 
this critical value depends on the value of the diffusivity 
(and also on the alpha coefficient) as it determines the growth 
rate \cite{karak}. Indeed, we do not know the exact value of 
diffusivity in the convection zone; we only have some idea of 
its order of magnitude estimate. Therefore, if we take a slightly 
higher (lower) value of diffusivity, then we can 
reproduce the Maunder minimum even at a higher (lower) critical value of $v_0$.

\begin{figure*}
\includegraphics[width=1.0\textwidth]{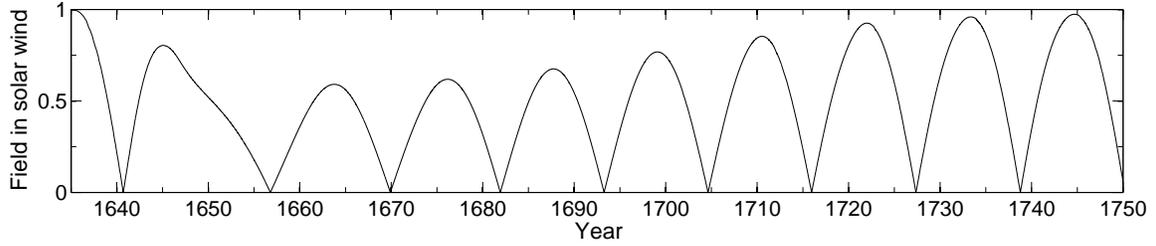}
\caption{Theoretically calculated average radial magnetic field (absolute
value in arbitrary units) in the solar wind.}
\label{wind}
\end{figure*}

Similar to Choudhuri \& Karak (2009), we calculate 
the magnetic field in the solar wind at a 
distance where it becomes radial. Fig.~\ref{wind} shows this as a 
function of time. This clearly shows the periodic 
behavior with a period slightly longer than the usual $11$~year period. This periodic variation explains the observed
oscillation in cosmogenic isotope data.

The cosmogenic isotope study suggests that there were around $27$ grand minima in last 11,000 years having 
different duration \cite{usos07}. Now in our simulation, the length of a grand minimum 
has been determined by the time for which the value of meridional circulation remains at a low value 
and the rate at which meridional circulation is increased after the grand minimum 
has started. In this way, we are able 
to reproduce any grand minimum of any length. 

\begin{figure*}
\centering
\includegraphics[width=1.00\textwidth]{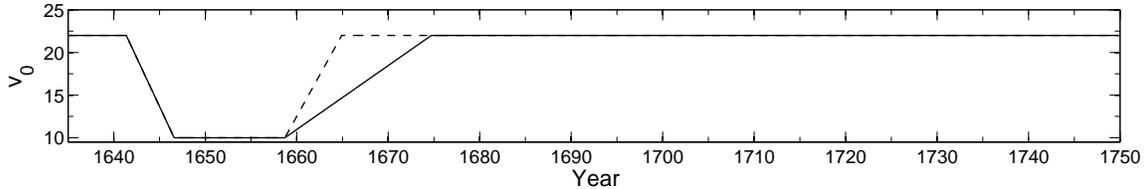}
\caption{Profile of amplitude of meridional circulation which produce 
the Maunder minimum similar to Fig.~\ref{mm}(b). The format is identical to Fig.~\ref{mm}(a).}
\label{mc}
\end{figure*}

We have no knowledge of how \mc\ varied during the 
Maunder minimum. In the above calculation, at the beginning of 
the Maunder minimum, we have decreased $v_0$ rapidly to a low 
value, whereas in the recovery phase, we increased it rather 
slowly. However, our results are completely 
independent of the way how we increase $v_0$ in the recovery phase. 
For example, the variation of $v_0$ in Fig.~\ref{mc} reproduces 
an exactly similar Maunder minimum what we have evinced in Fig.~\ref{mm}(b).
Here we underscore that if the \mc\ falls very slowly to a low
value, then we do not get the sudden fall of sunspot number at 
the beginning of the Maunder minimum. 
However, if we are to believe the observational results of 
Usoskin et al. (2000), then it says that at the beginning 
of the Maunder minimum, solar activity decreased abruptly, 
but built up gradually. This strongly suggests that 
the toroidal field decreased suddenly below a certain value 
which triggered the Maunder minimum and then 
it took sometime to regain its usual value. If this sudden 
fall of the toroidal field is due to the reduction of meridional circulation, 
then it probably indicates that the meridional circulation dropped rapidly to a very 
low value at the beginning. However, it does not indicate anything 
about the rate of recovery of meridional circulation.

Let us now discuss the physics of getting the Maunder minimum in this model. 
The meridional circulation affects the generation not 
only of the toroidal field but also that of the poloidal field (Wang \& Sheeley 2003; 
Baumann et al. 2004). Poloidal field generated at the surface through the 
decay of sunspots is advected by meridional circulation (and also diffuses) toward the 
pole where it cancels the opposite polarity field of the previous cycle to 
give rise to a large-scale polar field for the new cycle (Stenflo 1972; 
Wang et al. 1989; Dikpati et al. 2004). At the maximum 
phase of the solar cycle, lots of sunspot eruptions go on at the surface 
and at the same time, these sunspots decay in a short 
timescale to generate the poloidal field. Now at this time, 
if the amplitude of meridional circulation decreases strongly, 
then the poloidal field which is being generated at 
the surface at low latitude is not able to move toward 
the pole to cancel the opposite polarity field of the previous 
cycle rather a part of it diffuses away 
toward the opposite hemisphere to cancel out the opposite polarity field. 
Therefore, it is not able to generate a large amount of the polar field for 
the next cycle. Naturally, the decrease of meridional circulation 
causes a weaker polar field (see also Wang \& Sheeley 2003 and Baumann et al. 2004 
for details). This weaker polar field gives a weaker toroidal field (Choudhuri et al. 2007). 
In addition, according to the discussion of the previous section the lower value of 
meridional circulation means more time for the diffusive decay of the poloidal 
field during its transport through the convection zone, leading to less 
generation of the toroidal field in the tachocline. This weak toroidal field 
is insufficient to produce sunspot eruptions (D'Silva \& Choudhuri 1993; 
Fan et al. 1993; Caligari et al. 1995) and therefore triggers a grand minimum. 
However, when the amplitude of meridional circulation is increased again, 
the poloidal field does not get much time to diffuse and therefore the generation 
of the toroidal field is more. At some point, the toroidal field exceeds the 
critical value of producing sunspots eruption. In this way, it recovers from 
the grand minimum state.

The flux tube simulations (Choudhuri 1989; D'Silva \& Choudhuri 1993; Fan et al. 1993)
suggest that the initial magnetic field inside the flux tube has to be of order $10^5$ G
and the sunspot eruption takes place only when the toroidal field
reaches this value. In this model also, the sunspot eruptions take place
if the value of the toroidal field exceeds this critical value (Nandy \& Choudhuri 2001;
Chatterjee et al. 2004). During the Maunder minimum the toroidal field
at the base of the convection zone was below this value and there was 
no eruption. However, the weak solar cycles continued during this period. Note that in the
absence of sunspot eruptions, the usual Babcock--Leighton process does not work. However, the
toroidal field advects (and also diffuses) toward the upper part of the convection zone
by the upward meridional circulation. In this case, the $\alpha$ coefficient in our equation acts like
a traditional mean field $\alpha$ \cite{parker1955}. In modeling the distribution of a large-scale
solar magnetic field, Choudhuri \& Dikpati (1999) have shown that the poloidal field has
two sources---the Babcock--Leighton
mechanism and a $\alpha$-effect working on a weaker subsurface toroidal field. During
the Maunder minimum---when there was no sunspot---the dynamo continued due to the $\alpha$-effect
operating on the weaker toroidal field. Also the $\alpha$-effects operating in the
tachocline (Dikpati \& Gilman 2001; Bonanno et al. 2002, and references therein) 
or the alpha effect arising due to the buoyancy instability at the base of the 
convection zone \cite{ferri} may play a role in this episode. In this 
calculation, however, we do not consider these $\alpha$-effects.

\begin{figure}[!h]
\centering
\includegraphics[width=0.50\textwidth]{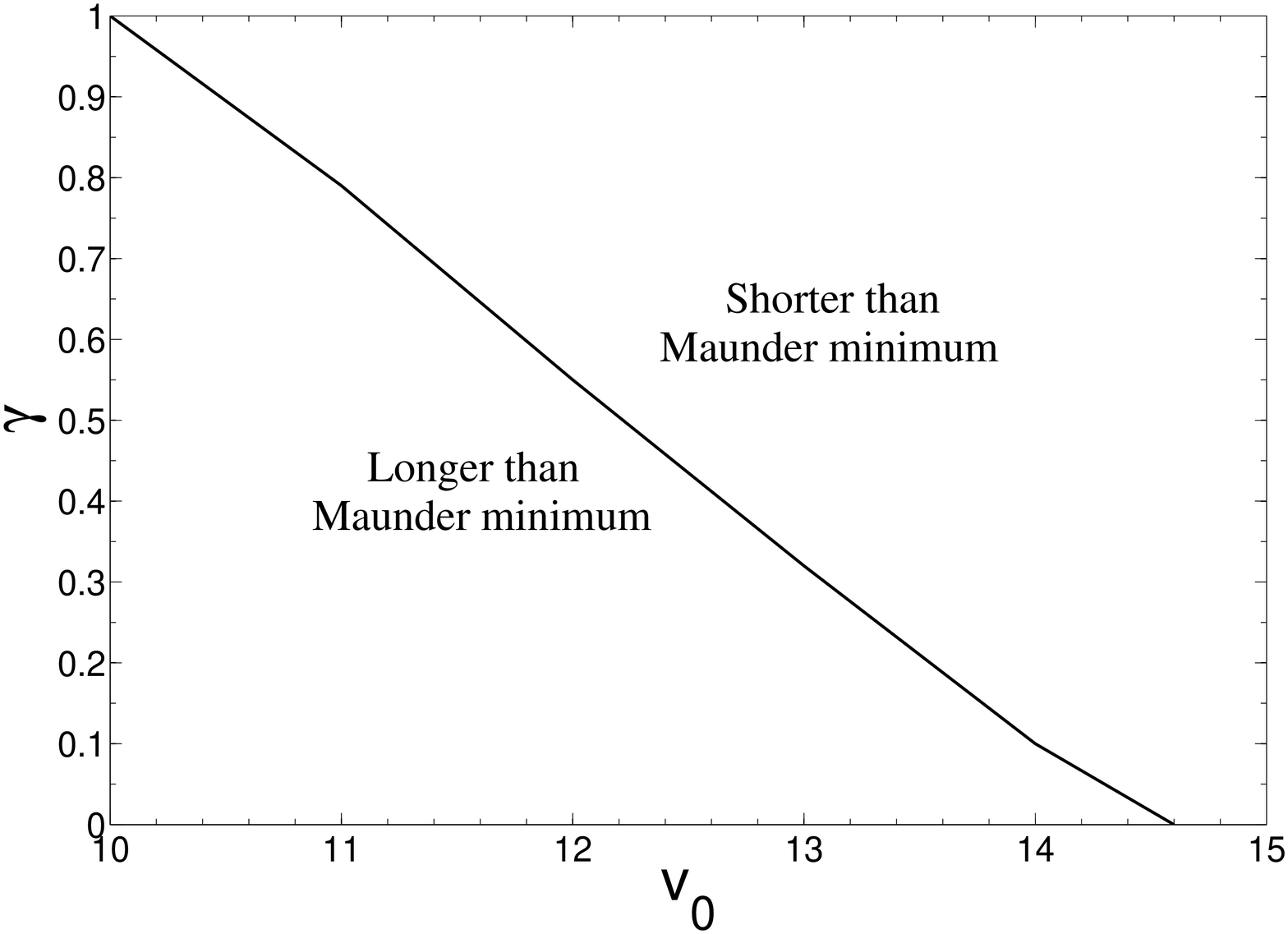}
\caption{Parameter space of amplitude of the \mc\ ($v_0$) and the poloidal field 
reduction factor ($\gamma$). The line shows the values of these parameters 
which give Maunder-like grand minima. 
}
\label{gv}
\end{figure}

Now we include the effect of fluctuations of the polar field along 
with the fluctuations of meridional circulation in this model. We have seen 
earlier that the strong decrease of $v_0$ can lead to a grand minimum. However, 
in that case, its value had to be reduced to a significantly low value which may be 
a bit surprising. We have mentioned that the polar field fluctuates randomly as its 
generation mechanism---the \bl\ process involves randomness. Therefore, it is 
important to include these fluctuations along with the fluctuations of 
meridional circulation in modeling grand minima. We find that if 
we decrease the polar field by a factor of $\gamma$, then even at a 
moderately lower value of $v_0$, we are able to reproduce Maunder-like grand minima. 
We have repeated our calculation at different values of $v_0$ and $\gamma$ and 
found that for each value of $v_0$ there is a corresponding value of 
$\gamma$ which can give a Maunder-like grand minimum. These values are shown 
by the line in Fig.~\ref{gv}. The values lower than these $\gamma$'s give 
grand minima longer than the Maunder minimum (like Sp\"orer minimum), whereas larger 
values give grand minima shorter than the Maunder minimum. These 
two regions are indicated in Fig.~\ref{gv}. We mention that 
Choudhuri \& Karak (2009) have reproduced 
the Maunder minimum just by decreasing the poloidal field to $0.2$ (actually $\gamma_N$ 
= $0.0$ and $\gamma_S$ = $0.4$) of its original value keeping $v_0$ unchanged. 
However, in the present calculations, we have to reduce $v_0$ to around $13.5$ m~s$^{-1}$ 
along with the poloidal field reduction to $0.2$. This is because in those 
calculations we had reduced the toroidal field slightly along with the 
change of the poloidal field to compensate for the overlap between cycles. 
Here we have not done this kind of questionable change of toroidal 
field. Additionally, here the values of $\alpha$ and $\eta_p$ 
are different than those calculations giving a different dynamo growth rate.
\begin{figure*}
\centering
\includegraphics[width=1.00\textwidth]{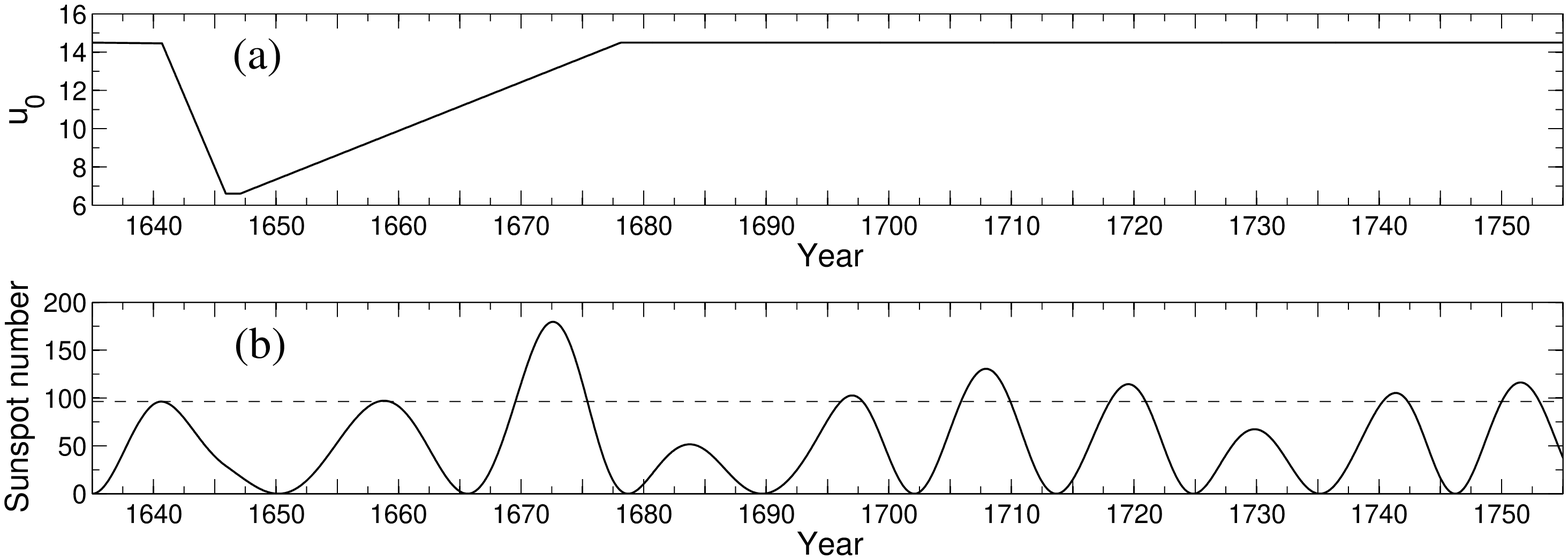}
\caption{Results from the low-diffusivity model. Upper panel shows the amplitude of 
meridional circulation ($u_0$ in m~s$^{-1}$) varied during Maunder minimum. The 
lower panel shows the variation of sunspot number ($B^2$) during this episode. The dashed 
line indicates the value of average sunspot number before Maunder minimum.}
\label{maundlow}
\end{figure*}

Now let us discuss the results of the low-diffusivity model briefly. In this 
model, lesser circulation speed means that the poloidal field 
gets more time to produce the toroidal field in the tachocline. Thus smaller $v_0$ gives 
more toroidal field and finally stronger cycle. Consequently, if the dynamo is in 
the advection-dominated regime, it does not produce any grand minimum rather it produces 
stronger cycles. This is exactly what we find in our simulation. The results are 
shown in Fig.~\ref{maundlow}. In this figure, panel (a) shows the variation of $u_0$ during
the Maunder minimum episode, whereas panel (b) shows the theoretical sunspot number 
with time. We see that the sunspot number has rather increased when the 
meridional circulation is decreased rapidly from its 
average value ($14.5$ m~s$^{-1}$) to a very low value ($6.6$ m~s$^{-1}$). However, 
for one cycle (around 1680--1690 and also around 1725--1735) after two cycles 
of the strong reduction of $u_0$ we get a reduction of sunspot number. This is due 
to the decrease of the polar field during the low value of $v_0$ as explained earlier. 
It may be noted that in this model, the decrease 
of the polar field affects the sunspot cycle of two cycles later (see figure 9 of 
Jiang et al. 2007). We may point out that if we increase $u_0$ to 
a very high value for several years, then we may get a Maunder-like minimum 
in this model. However, in this case, the cycle periods will be 
unrealistically short and no observation during the Maunder minimum validates this.

\section{CONCLUSION}
We have studied the importance of meridional circulation on the period and the 
amplitude of solar cycle and also on the Maunder-like grand minimum in two 
different regimes of a flux transport dynamo model, namely, advection-dominated 
and diffusion-dominated. First, we have approximately modeled the periods of 
the last $23$ solar cycles by varying $v_0$ alone. From this study, we get some 
idea about both the amplitude variation and the time scale of the \mc\ over 
the last few centuries. Moreover, we have seen that
when we match the periods of these cycles by varying $v_0$ in 
the high-diffusivity model, most of the cycle amplitudes also get 
modeled up to some extent. Therefore, we conclude that 
a major part of the fluctuations of solar cycle amplitude 
may come from the \mc\ fluctuations. In the low-diffusivity 
model, the cycle amplitudes do not get modeled at all 
when we repeat the same analysis.

We have also shown the possibility of producing (or not producing) 
a Maunder-like minimum in two models by varying the 
meridional circulation speed. To do this, we have assumed that 
at the beginning of the Maunder minimum the amplitude of meridional circulation 
largely reduced and then after a few years it increased again. 
It may be noted that this is not an ad hoc assumption. There are several 
independent arguments which support our assumption. First, 
Wang \& Sheeley (2003) used the flux transport model to simulate the evolution of the Sun's magnetic 
dipole moment, polar fields, and open flux under Maunder minimum conditions 
and suggested that the poleward surface flow speed was reduced from $20$ m~s$^{-1}$ to $10$ m~s$^{-1}$ during that time. 
Second, the flux transport dynamo model predicts 
an inverse correlation between the meridional circulation speed and the cycle period (Wang et al. 1991; 
Dikpati \& Charbonneau 1999). On the other hand, from the study of $^{14}$C data during the Maunder minimum, 
Miyahara et al. (2004) (also Miyahara et al. 2010) 
reported that the periods of the solar cycle were longer 
compared to the usual 11~year period. 
Now, if the flow speed determines the solar cycle 
period, then it indicates that the meridional circulation speed during 
the Maunder minimum was lower than the usual value. 
Third, while simulating Sun's large-scale magnetic 
field during cycles 13--22, Wang et al. (2002) have shown that the regular polarity reversal 
is obtained only if the flow speed is assumed to be correlated 
with the cycle amplitude (see also Hathaway et al. 2003). 
On the other hand, Yeates et al. (2008) have shown that the cycle amplitude is positively correlated 
with the speed of meridional circulation in the diffusion-dominated regime. 
Now, both Beer et al. (1998) and Miyahara et al. (2004) 
found that the amplitudes of solar cycle were weaker during 
the Maunder minimum. Therefore, if flow speed determines the cycle amplitude, then it 
probably indicates that the meridional circulation during
the Maunder minimum was weaker than the usual one. Last, using the low order 
dynamo model, Passos \& Lopes (2009) suggest that the stochastic fluctuations 
in the $\alpha$-effect cannot trigger a grand minimum rather they argued that the 
decrease in the amplitude of the meridional flow can do so.

We have no idea why the meridional circulation dropped to a very low value. 
Therefore, our assumption may be a bit questionable 
at the present understanding. However, this assumption enables us to reproduce most of 
the important features of the Maunder minimum remarkably well. It may be noted that 
to produce a Maunder-like minimum in our model, we did not have to decrease $v_0$ 
to an unrealistically low value ($10$~m~s$^{-1}$ is required here). 

We emphasize that we can reproduce these results 
only if the dynamo is in the diffusion-dominated regime. 
However, recently several independent works on the parity 
(Chatterjee et al. 2004; Hotta \& Yokoyama 2010b), 
the hemispheric coupling (Chatterjee \& Choudhuri 2006; Goel \& Choudhuri 2009), 
the correlation between the polar field at the minimum and the sunspot number in
the next cycle (Jiang et al. 2007), 
the correct value of the polar field (Hotta \& Yokoyama 2010a), the correlation between the 
flow speed and the cycle amplitude (Wang et al. 2002; Hathaway et al. 2003; 
Yeates et al. 2008) and the Waldmeier effect (Karak \& Choudhuri 2010b) 
suggest that the solar dynamo is working in the diffusion-dominated regime and 
not in the advection-dominated regime. In addition, from the low-order time-delay 
dynamo model, Wilmot-Smith et al. (2006) have shown that the irregular activity 
and the Maunder-like events are more readily excited in the diffusion-dominated 
regime and not in the advection-dominated regime.

However, we are not pointing out that the decrease of meridional circulation is the 
``only'' possibility to produce grand minima, but the abrupt decrease of the polar field 
due to fluctuations in the Babcock--Leighton process may be a valid possibility 
\cite{karak}. As the measurement of either the meridional circulation or the polar 
field during the Maunder minimum is unavailable, we do not know 
which possibility is correct. In this paper, we have adequately shown 
that a large fluctuation of meridional circulation alone can lead to a 
grand minimum.

\noindent{\it Acknowledgments:}
It is my pleasure to thank Arnab Rai Choudhuri for encouraging 
me to write this paper and also for helping me during the preparation 
of the paper by giving constructive suggestions. I also thank 
Dibyendu Nandy, Piyali Chatterjee, Jie Jiang, and Dipankar Banerjee for the fruitful discussions. 
Financial supports from J. C. Bose Fellowship of Arnab Rai Choudhuri (project no. SR/S2/JCB-61/2009) 
and CSIR, India are acknowledged.

\end{document}